\documentclass[pre,aps,10pt,twocolumn,floatfix]{revtex4-1}

\usepackage{graphicx}
\usepackage{amsmath}
\usepackage{amsfonts}

\renewcommand{\vec}[1]{{\bf{#1}}}
\renewcommand{\rm}[1]{{\mathrm{#1}}}
\newcommand{\LP}{LP}
\def\l{\left}
\def\r{\right}

\begin{document}

\title{Sparse phonon modes of a limit-periodic structure}

\author{Catherine Marcoux}
\affiliation{Physics {D}epartment, Duke University, Durham, NC 27708}

\author{Joshua E.~S.~Socolar}
\email[]{socolar@phy.duke.edu}
\affiliation{Physics {D}epartment, Duke University, Durham, NC 27708}

\date{\today}

\begin{abstract}
Limit-periodic structures are well ordered but nonperiodic, and hence have nontrivial vibrational modes.  We study a ball and spring model with a limit-periodic pattern of spring stiffnesses and identify a set of extended modes with arbitrarily low participation ratios, a situation that appears to be unique to limit-periodic systems.  The balls that oscillate with large amplitude in these modes live on periodic nets with arbitrarily large lattice constants.  By studying periodic approximants to the limit-periodic structure, we present numerical evidence for the existence of such modes, and we give a heuristic explanation of their structure.
\end{abstract}


\maketitle


\section{Introduction}
\label{sec:intro}

Nonperiodic structures are known to support vibrational modes that differ markedly from the Bloch waves of infinite periodic crystals.  Well studied examples include the localized modes of disordered systems~\cite{sievers_RMP47} or floppy materials~\cite{during_SoftMatt9}, the critical modes of quasiperiodic systems, which exhibit power-law decays~\cite{janssen_RMP69}, and the topologically protected modes associated with boundaries or line defects in isostatic lattices~\cite{lubensky_PNAS109, vitelli_NatPhys11} or mechanical models with broken time-reversal symmetry~\cite{prodan_PRL103} or chiral couplings~\cite{pal_arXiv:1511.07507}. An interesting feature of such systems is the possibility of the localization of mechanical energy on low dimensional structures.~\cite{zhang_PRB75, liu_APL95, laude_APL84, witten_RMP79}. Typically, the localized modes occur near defects in crystals and at surfaces~\cite{yodh_PRE88, sigalas_JAP84}.

Limit-periodic (\LP) structures occupy a conceptual space in between periodic crystals and quasicrystals or disordered systems.  Like crystals and quasicrystals, they are homogeneous in the sense that every local region in them is repeated with nonzero density, and they are translationally ordered, having diffraction patterns that consist entirely of Bragg peaks.  Unlike crystals, however, there is no smallest wavenumber in the diffraction pattern; the set of Bragg peaks is dense.  But unlike quasicrystals, the point group symmetry of a \LP\ structure is compatible with periodicity, and the structure can be described as a union of periodic structures with ever increasing lattice constants~\cite{socolar-taylor_JCombTh11}.  It is thus natural to ask whether they support modes with novel spatial structures.  In particular, one might wonder if the \LP\ structure could support modes with low participation ratios.

Though no naturally occurring \LP\ structures have been discovered, a recent result in tiling theory shows that local interactions among tiles that are identical up to reflection symmetry can favor the production of two- or three-dimensional hexagonal \LP\ structures~\cite{socolar-taylor_JCombTh11,marcoux-socolar_PRE90}.  It has also been shown in simulations that a collection of identical achiral units with only nearest neighbor interactions can spontaneously form a hexagonal limit-periodic structure when slowly cooled~\cite{byington-socolar_PRL108, marcoux-socolar_PRE90}. With recent advances in colloidal particle synthesis, the fabrication of particles with the necessary interactions for formation of the \LP\ structure seems experimentally feasible~\cite{wang-pine_Nature, yi-sacanna_JPhysCM, feng-chaikin_AdvMat, petsev_PRL113}.  The possibility of creating a \LP\ phase motivates us to explore the physical properties associated with its unique translational symmetries. 

Here we study the spectrum of a \LP\ structure inspired by the Taylor-Socolar tiling~\cite{socolar-taylor_JCombTh11, socolar-tayor_MathInt12}.  Our system consists of identical point masses placed on the sites of a triangular lattice and connected by springs on all of the nearest neighbor bonds.  The springs are assigned one of two possible stiffnesses, where the pattern of assignments is \LP.  To study the vibrational spectrum, we construct a hierarchy of periodic approximant models and use standard techniques to calculate their phonon modes.  We observe that certain modes with low participation ratios remain unchanged as the lattice constant of the approximant increases and that at each new scale additional modes arise with even lower participation ratios.  Though these modes are extended, and indeed are perfectly periodic, the particles that oscillate with large amplitude are confined to sparse networks of 1D chains.  We also find that these modes are not destroyed by vacancies or by small amounts of disorder in the spring constants. 

The rest of the paper is ordered as follows.  In Section~\ref{sec:limit-periodic}, we describe the \LP\ structure of interest, its periodic approximants, and the corresponding ball-and-spring models. Section~\ref{sec:computational_methods} presents the methods used to compute the spectra of the approximants. Section~\ref{sec:identify_modes} shows how modes of the infinite \LP\ structure are identified and describes their structure.  Section~\ref{sec:origin} presents an analysis of the origin of the modes of interest.


\section{A limit-periodic ball-and-spring model}
\label{sec:limit-periodic}

\begin{figure}[h]
    \centering
    \includegraphics[width=0.4\textwidth]{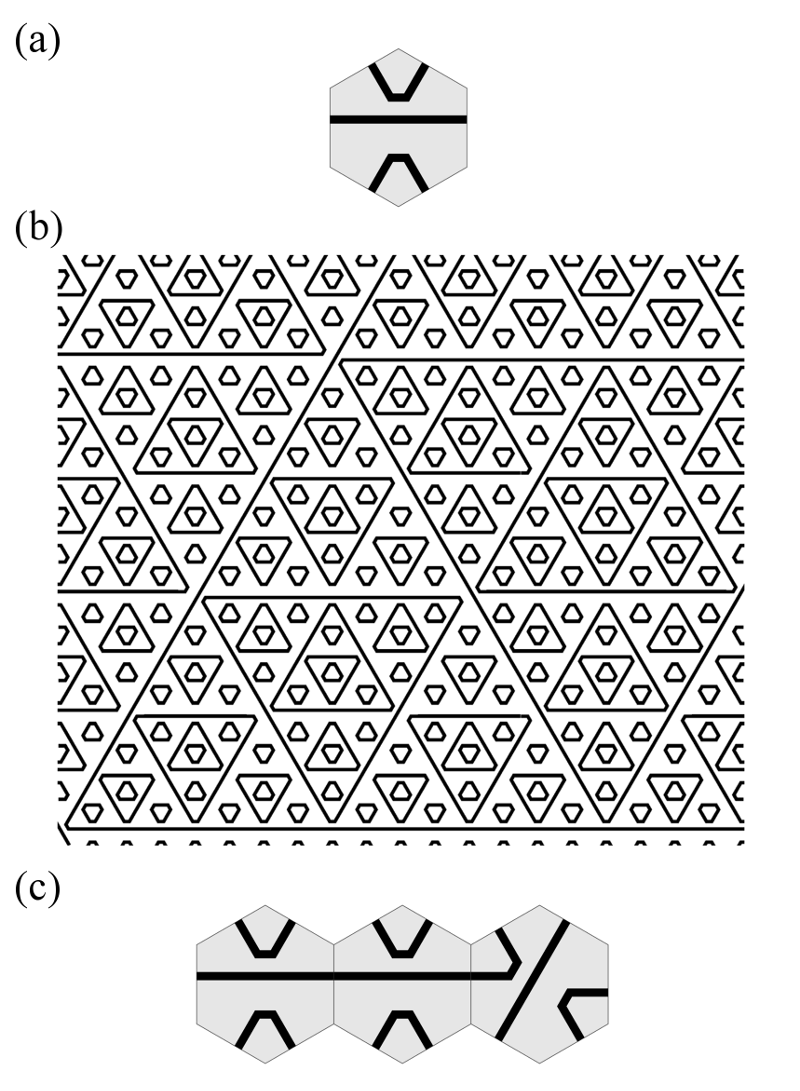}
    \caption{(a) The tile that can be used to create the \LP\ structure. (b) A section of the \LP\ structure. (c) An example of a local configuration of tiles.}
    \label{fig:limit-periodic}
\end{figure}

The \LP\ pattern studied here is formed from a dense packing of a single type of decorated tile: the hexagon with black stripes shown in Fig.~\ref{fig:limit-periodic}(a). The 
tiles are arranged on a triangular lattice and oriented as shown in Fig.~\ref{fig:limit-periodic}(b). The structure is completely homogeneous in the sense that it consists of a uniform density of identical tiles.
We note that in a statistical mechanical lattice model of this system this \LP\ structure forms spontaneously in a slow quench from a state of disordered tile orientations~\cite{marcoux-socolar_PRE90}.

\begin{figure}[h]
    \begin{center}
    \includegraphics[width=0.4\textwidth]{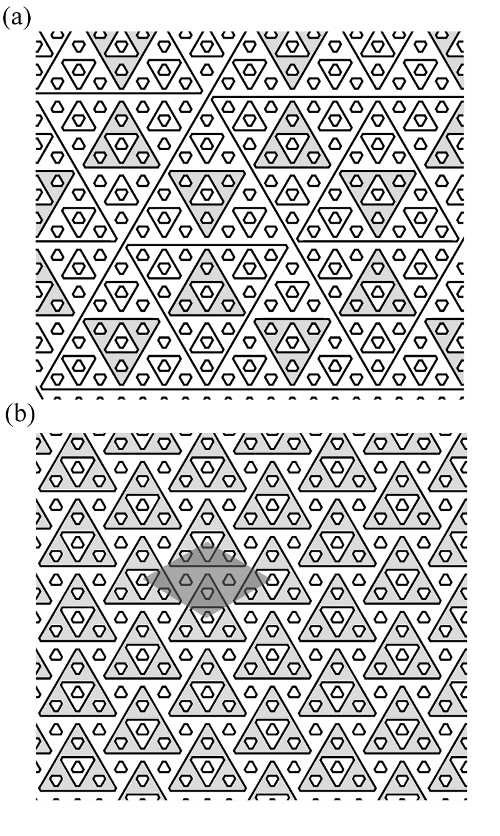}
    \caption{(a) The \LP\ structure. (b) The 3-periodic structure with the unit-cell colored dark gray.  The level-3 triangles are filled to highlight the difference in the pattern of level-3 triangles within the structures shown in (a) and (b). }
    \label{fig:periodic}
	\end{center}
\end{figure}

As mentioned above, a \LP\ structure consists of a union of periodic crystals with ever larger lattice constants.  In the present case,
 each set of triangles of a given size forms a crystal, with the centers of the triangles at the vertices of a honeycomb. Fig.~\ref{fig:limit-periodic}(c) shows the way that neighboring tiles join to form the edges and corners of all but the smallest triangles.  
The number of tiles that contribute decorations to form a triangle is $3\times2^{n-1}$, where $n$ is any positive integer.  Three of these tiles create the corners, while the rest form the edges.  We refer to a triangle with a given $n$ as a {\em level-$n$} triangle, and we refer to the entire pattern of such triangles as level~$n$.  The shading in Fig.~\ref{fig:periodic}(a) highlights the level-3 triangles.  Note that the level-$n$ pattern has exact 6-fold rotational symmetry for all $n$.
The \LP\ structure is 6-fold symmetric in the sense that every bounded configuration that appears is repeated with equal density in all six orientations corresponding to rotations by $\pi/3$.

\begin{figure}
  \centering
  \includegraphics[width=0.41\textwidth]{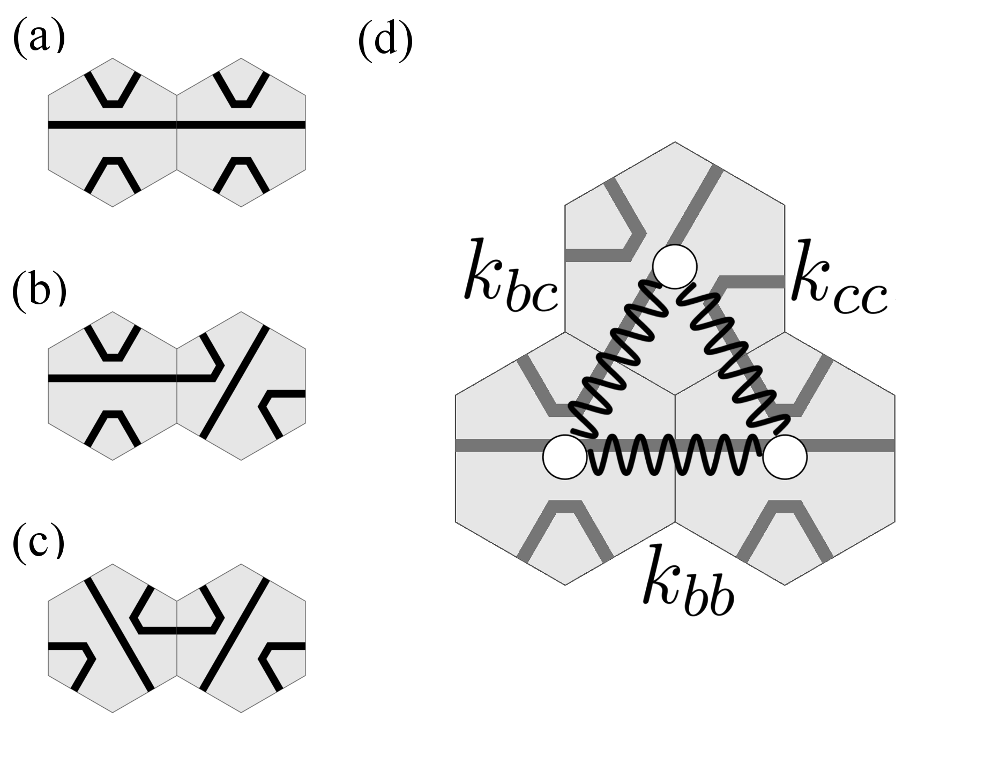}
  \caption{Examples of the three types of bonds in the \LP\ structure: (a) a bar-bar bond $(bb)$, (b) a bar-corner bond $(bc)$, and (c) a corner-corner bond $(cc)$. (d) A depiction of the ball and spring model. The coupling strength of a spring connecting nearest neighbor masses is determined by the type of bond.}
  \label{fig:ball_and_spring}
\end{figure}

To develop a physically plausible ball-and-spring model, we place a point mass at the center of each hexagonal tile and connect nearest neighbors with springs, where the stiffness of a given spring is determined by the configuration of black stripe decorations across the boundary between the two tiles.
In the \LP\ structure, there are three types of nearest neighbor bonds: 
bar-bar ($bb$), bar-corner ($bc$), and corner-corner ($cc$), as illustrated in
Fig.~\ref{fig:ball_and_spring}(a)-(c). 
Spring stiffnesses $k_{bb}$, $k_{bc}$, and $k_{cc}$ are assigned to the $bb$, $bc$, and $cc$ bonds, respectively, as shown in Fig.~\ref{fig:ball_and_spring}(d).  Note that for $n\ge 3$, level $n$ is formed by $k_{bb}$ chains of length $2^{n-1}-1$ coupled through $k_{bc}$ bonds at the triangle corners. 

The relative densities of the different bond types
is set by the level-1 structure.  First note that the total number of bonds is $3$ per tile.  All $cc$ bonds are formed by the tiles that create the level-1 triangles.  In the \LP\ pattern, $3/4$ of the tiles contribute four corner bonds each to level~1.  As each $cc$ bond is counted twice in this manner, the number of $cc$ bonds is $3/2$ per tile.  The corners on each of the remaining $1/4$ of the tiles form $bc$ bonds, again with each tile contributing to four corner bonds.  In this case, each bond is only counted once, so the number of $bc$ bonds is $1$ per tile.  The remaining bonds, $1/2$ per tile, must be $bb$ bonds.  Thus in any pattern in which level 1 is the honeycomb of the \LP\ structure, 
$1/2$ of the bonds are $cc$, $1/3$ are $bc$, and $1/6$ are $bb$.  All of the models considered below, including the periodic approximants, have this property, making the average spring stiffness the same in all cases.

There exist periodic tilings of the decorated hexagon of Fig.~\ref{fig:limit-periodic}(a) that contain elements of the \LP\ structure. (Note that the next nearest neighbor interactions required to force aperiodicity of the Taylor-Socolar tile are not enforced by this decoration.)
For present purposes, we construct a series of periodic approximants of the type shown in Fig.~\ref{fig:periodic}(b) and refer to them as $n$-periodic. In an $n$-periodic structure, the largest triangles are level-$n$.  A crucial feature of these approximants is that levels 1 through $(n-2)$ are identical to their counterparts in the \LP\ structure. The $n$-periodic structure has 3-fold rotational symmetry and a unit cell consisting of $3\times4^{n-2}$ tiles. For $n\ge 3$, the level-1 structure is identical to that of the \LP\ pattern, so the ratio of densities of the bond types is also the same.
For completeness we note that there does exist a 2-periodic structure in which $1/3$ of the bonds are $cc$, $2/3$ are $bc$, and there are no $bb$ bonds, but it is not relevant for present purposes.

\begin{figure}[h]
    \centering
    \includegraphics[width=0.5\textwidth]{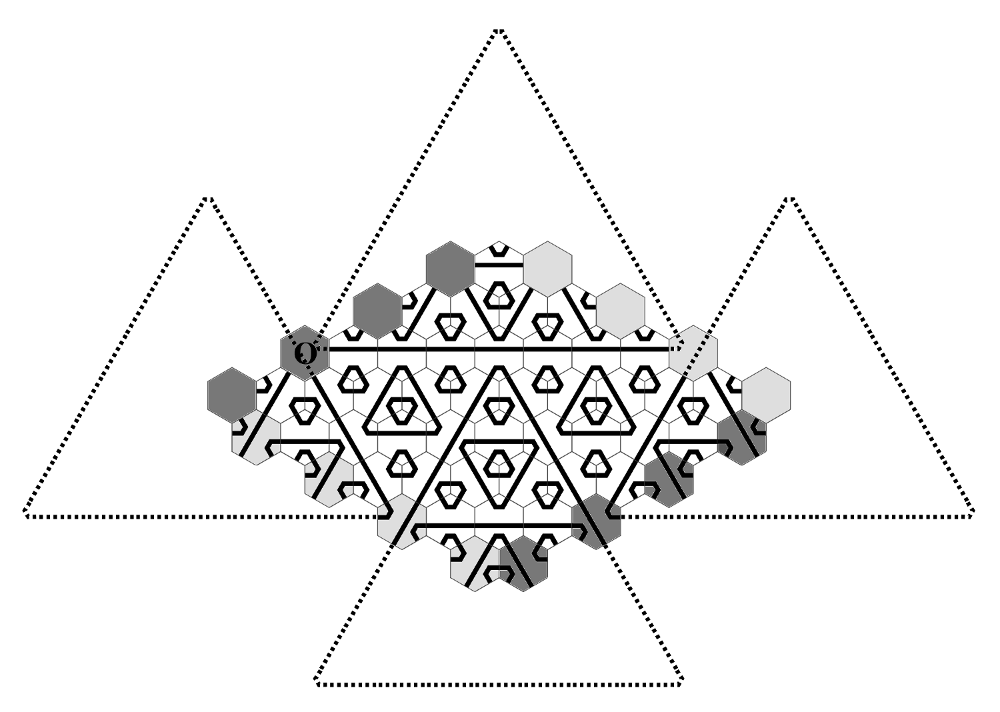}
    \caption{The unit cell of the 4-periodic structure. The dashed lines are drawn to show how the largest triangles are formed by the unit cell. The shades of the tiles indicate edges that are equivalent when periodic boundary conditions are applied. }
    \label{fig:4-periodic_unit_cell}
\end{figure}

Fig.~\ref{fig:4-periodic_unit_cell} shows the 4-periodic structure.  In this structure levels~3 and~4 do {\em not} have the honeycomb pattern characteristic of the \LP\ structure, while levels~1 and~2 do.  In general, the $n$-periodic approximants with larger $n$ have the same structure, with the dotted triangles in the figure indicating level-$n$ triangles and all levels below and including $n-2$ having the same structure as they do in the \LP\ case.

In the ball-and-spring models studied here, all the balls are taken to have the same mass $\mu$, in accordance with the fact that the tiles are all identical.  The coupling strengths $k_{ij}$ are assigned as described above, and all springs are taken to have an unstressed length equal to the lattice constant $a$.  An algebraic formula specifying $k_{ij}$ at a given location in the \LP\ structure is given in the Appendix.


\section{Computational methods}
\label{sec:computational_methods}

To determine whether low participation ratio modes exist in the \LP\ structure, we study the periodic approximants and extrapolate our results. The bulk of the numerical analysis is done using the 7-periodic structure, but specific modes of the 5-, 6-, and 8-periodic structures were also calculated. 

\begin{figure*}
  \centering
  \includegraphics[width=\textwidth]{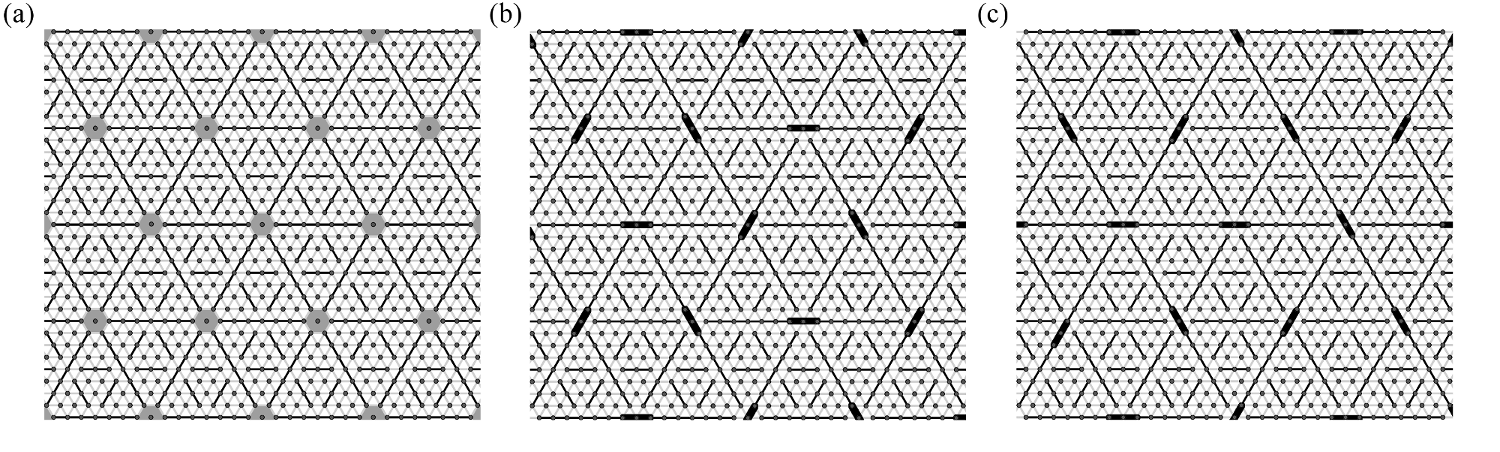}
  \caption{Coupling patterns for $n$-periodic approximants with $n\ge5$.  Black (light gray) lines represent bonds with coupling constant $k_{bb}$ ($0.6k_{bb}$). (a) Coupling pattern common to all approximants and the \LP\ structure.  The different cases correspond to particular patterns of springs within the dark gray hexagons: (b) the 5-periodic couplings.  (c) the \LP\ coupling pattern.  The $k_{bb}$ bonds in (b) and (c) are bolded to highlight the difference between them.}
  \label{fig:5-periodic_removed}
\end{figure*}

Following standard practice for a lattice with a basis~\cite{ashcroft_mermin_solidstate76}, we let $\vec u_{i}(\vec R,t)$ denote the displacement of the particle at equilibrium position {\vec R}. The index $i$ specifies which element of the basis corresponds to position $\vec R$. For a normal mode with wavevector $\vec q$ and frequency $\omega$ we have 
\begin{equation} 
\vec u_{i}(\vec R, t)  =  \Re\l[ \boldsymbol \epsilon_{i} e^{i(\vec q \cdot \vec R - \omega t)}\r] = (\mathrm{u}_{i\mathrm{x}}(\vec R, t), \mathrm{u}_{i\mathrm{y}}(\vec R, t)) \,,
\label{eq:ui}
\end{equation}
where $\Re$ denotes the real part and $\boldsymbol \epsilon_{i}$ is a polarization vector that is the same for the particle in each unit cell corresponding to basis element $i$. The $\boldsymbol \epsilon_i$'s are normalized such that $\sum_{i} \boldsymbol \epsilon_i \cdot \boldsymbol \epsilon_i = 1$, where the sum runs over the sites in one unit cell. Defining
\begin{equation}
	\vec f_{i} = \frac{1}{\mu}\sum_{j=1}^6 k_{ij}  \l[ \l( \boldsymbol \epsilon_{i} -\boldsymbol \epsilon_{j} e^{i\vec q\cdot  \vec  n_{ij}} \r)\cdot  \hat{\vec  n}_{ij} \r] \hat{\vec  n}_{ij} = (\rm f_{i\mathrm{x}}, \rm f_{i\mathrm{y}})\,,
\end{equation}
where $\vec n_{ij} = \vec r_{j} - \vec r_{i}$, $\vec r_{i}$ is the position of particle $i$, and $k_{ij}$ is the coupling strength of the bond between particle $i$ and nearest neighbor $j$, and the vectors
\begin{align}
\vec F &= (\rm f_{\mathrm{1x}}, \rm f_{\mathrm{1y}},\rm  f_{\mathrm{2x}},\rm  f_{\mathrm{2y}},...,\rm f_{\mathrm{Nx}}, \rm f_{\mathrm{Ny}})\\
\vec E & = ( \epsilon_{\mathrm{1x}}, \epsilon_{\mathrm{1y}}, \epsilon_{\mathrm{2x}}, \epsilon_{\mathrm{2y}}, ... ,  \epsilon_{\mathrm{Nx}}, \epsilon_{\mathrm{Ny}})
\end{align}
 one constructs the dynamical matrix $\vec D(\vec q)$ with elements
\begin{equation}
\vec D_{n,m}(\vec q) = \frac{ \partial  \mathrm{F}_{n}}{\partial  \mathrm{E}_{m}}\,,
\end{equation}
where $n$ and $m$ integers between 1 and $2N$. The normal modes and their frequencies are determined by the eigenvalue equation
\begin{equation}
\l( \vec D(\vec q) -\omega^2 \mathbb{I} \r) \vec E = 0\,.
\end{equation}
After constructing the dynamical matrix corresponding to the proper assignment of coupling strengths $k_{ij}$, we use standard Mathematica functions to solve for $\omega$ and $\vec E$.

We report results for coupling strengths
\begin{equation}
k_{cc} = k_{bc} = \alpha k_{bb}\,,
\label{eq:spring_constants}
\end{equation}
with $\alpha < 1$. For purposes of illustration, we choose $\alpha = 0.6$. The qualitative features do not depend on the particular values of the coupling strengths as long as $k_{cc}$ and $k_{bc}$ are both less than $k_{bb}$. Without loss of generality we set $k_{bb}/\mu = 1$ and $a = 1$.

The limit of interest is a structure that has no periodically repeated unit cell and hence requires an infinite set of polarization vectors for each mode.  For any $n$-periodic approximant, the number of polarization vectors required is $3\times4^{n-2}$.  When $n$ is increased by one, the Brillouin zone shrinks in area by a factor of 4, and the number of modes at any given wavenumber within the new Brillouin zone grows by that same factor.  In the limit of infinite $n$ all of the modes are formally $\vec{q}=0$ modes.  To explore the structure of these modes, we consider only the $\vec{q}=0$ modes of each approximant, which turn out to have features that allow for extrapolation to the full \LP\ system.


\section{Modes of the limit-periodic structure}
\label{sec:identify_modes}

Given the homogeneity of the structure, we expect the low frequency modes of all of the approximants to be small perturbations of ordinary plane waves.  We have confirmed that the sound speed is isotropic and corresponds to that of a triangular lattice with coupling constant $0.659k_{bb}$, which is roughly equal to the weighted average of the coupling strengths in the unit cell $\langle k \rangle = 0.667$.  The more interesting portion of the spectrum contains the high frequency modes, which are sensitive to variation of couplings on all scales.

The identification of modes of the \LP\ structure rests on the surprising observation that certain modes are simultaneously normal modes of the $n$-periodic and the \LP\ structures. To see how this may be possible, consider the pattern of bond strengths depicted in Fig.~\ref{fig:5-periodic_removed}(a). If there is a mode in which all of the springs within the shaded hexagons remain unstressed to first order, that mode is entirely insensitive to the pattern of coupling strengths within each hexagon. In particular, the coupling strengths can be chosen to create the 5-periodic structure shown in Fig.~\ref{fig:5-periodic_removed}(b) or, alternatively, to create the \LP\ structure shown in Fig.~\ref{fig:5-periodic_removed}(c), or indeed to create any $n$-periodic approximant with $n>5$.  
Because the pattern in Fig.~\ref{fig:5-periodic_removed}(a) has 6-fold symmetry about each shaded hexagon, there can be modes that exhibit 3-fold or 6-fold symmetry about these points as well, as long as they do not involve any stretching of the bonds within the shaded hexagons.  If such a mode does exist, then it is a mode of any of the approximants of higher order. We find numerically that there are many such modes.

\begin{figure*}
  \centering
  \includegraphics[width=\textwidth]{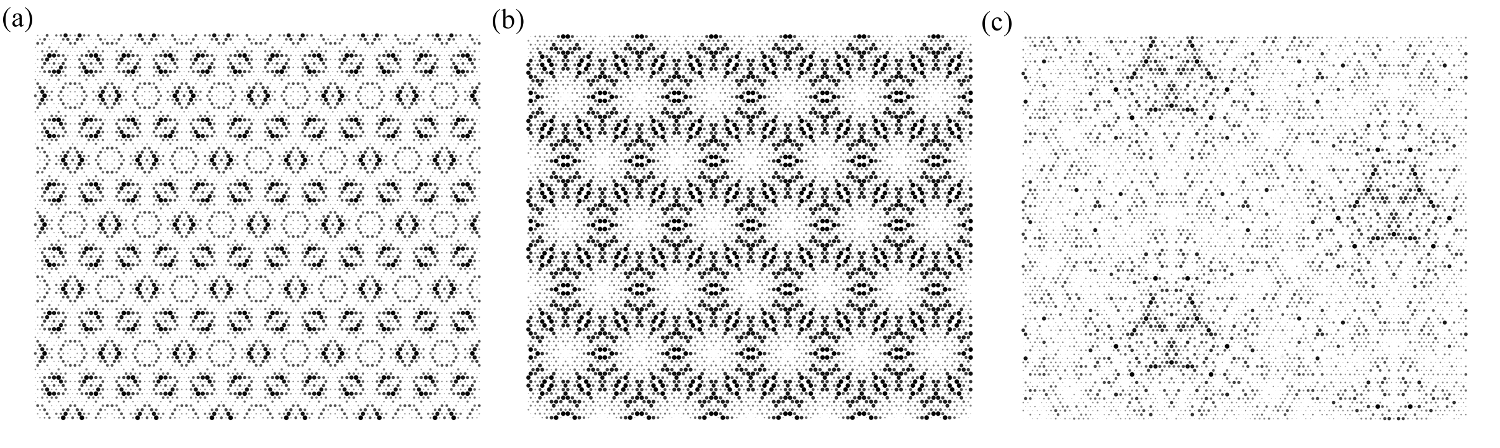}
  \caption{ A selection of $\vec q = 0$ modes of the 7-periodic structure with typical participation ratios. Dot size indicates the amplitude of oscillation of the corresponding mass. (a,b) Modes that are also modes of the \LP\ structure.  (c) A mode that reflects the unique periodicity of the 7-periodic approximant. The frequencies and participation ratios of the modes are: (a)  $\omega = 1.772$, $p = 0.611$; (b)  $\omega = 1.378$, $p = 0.725$; and (c) $\omega =1.225$, $p = 0.637$.}
  \label{fig:long_wavelength}
\end{figure*}

To identify modes of particular interest, we calculate for each mode a participation ratio $p$ defined as~\cite{hibbins_butler_JPhysC1970, schober_PRL66} 
\begin{equation}
p = \l ( N \sum_{i=1}^N |\boldsymbol \epsilon_i|^4 \r)^{-1}\,,
\label{eq:participation_ratio}
\end{equation}
where the normalization of $\boldsymbol \epsilon_i$ yields $p=1$ if $|\boldsymbol \epsilon_i|$ is the same for all $i$.  We find that most modes of the 7-periodic structure have $p\sim 0.6$.  Some typical examples are shown in Fig.~\ref{fig:long_wavelength}.   

\begin{figure*}
  \centering
  \includegraphics[width=\textwidth]{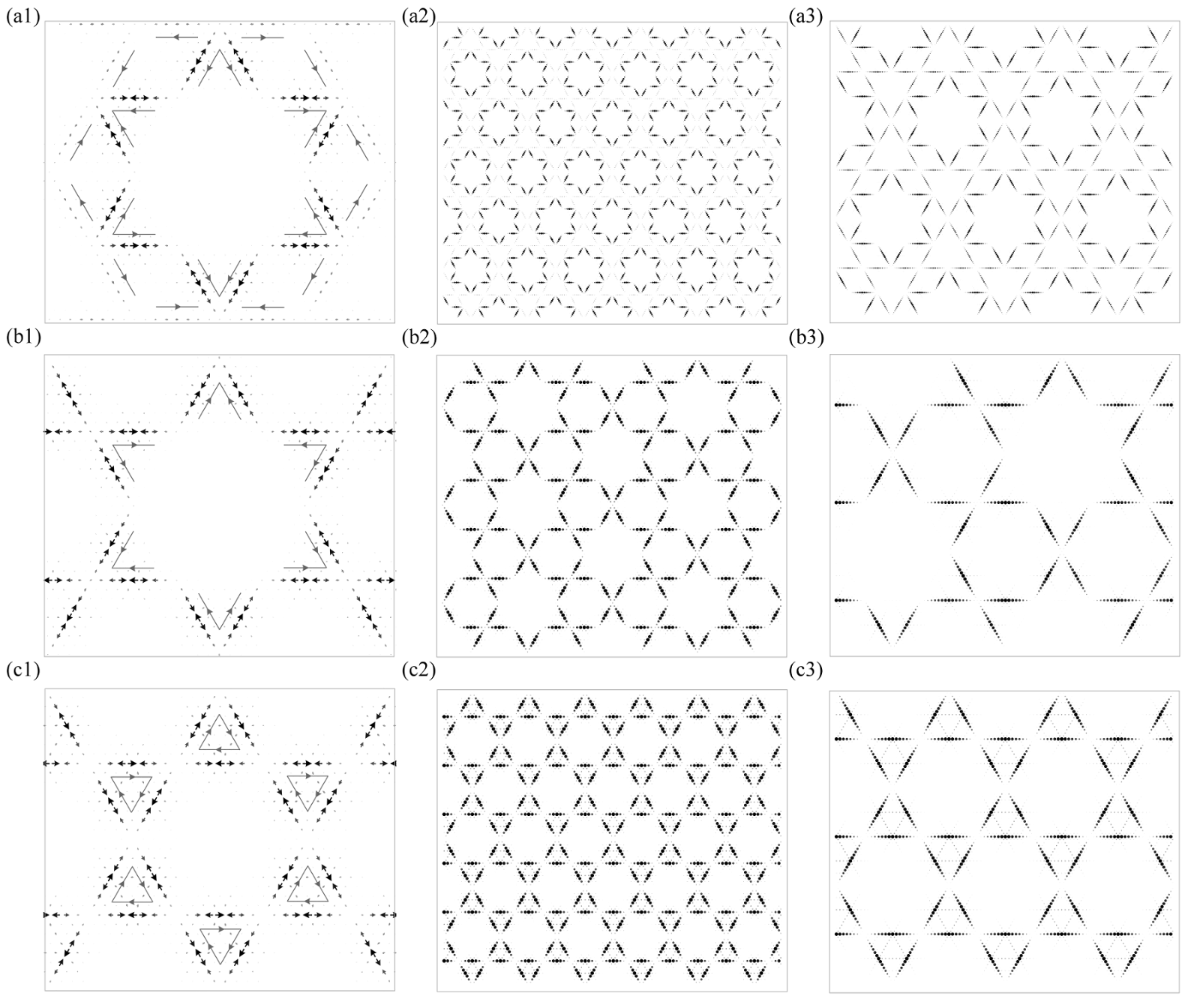}
  \vspace{-12pt}
  \caption{Low participation ratio modes.  Column~(1): Snapshots of unit cells. Black arrows indicate the displacement of the masses at a given time.  Gray arrows indicate the symmetry of the instantaneous pattern of displacements. Columns~(2) and~(3):  Members of the hierarchy corresponding to column~1. Black dot sizes indicate the amplitude of the polarization vector at each lattice site. Row (a): Two-level modes of the 8-periodic structure.  (a2): Levels~4 and~5 are excited; $\omega = 2.161$; $p = 0.138$. (a3): Levels~5 and~6 are excited; $\omega = 2.187$; $p = 0.103$. Row (b):  Single-level modes of the 7-periodic structure. (b2): Level~4; $\omega = 2.163$; $p = 0.144$.  (b3): Level~5; $\omega = 2.188$; $p = 0.075$. Row (c): Modes of the 7-periodic structure in which all of the edges of a single level are excited.  (c2): Level~4; $\omega = 2.162$; $p = 0.208$.  (c3): Level~5; $\omega = 2.189$; $p = 0.118$.}
  \label{fig:panel_of_modes}
\end{figure*}

The $n$-periodic approximant supports modes with very low participation ratios, many of which have the 3-fold or 6-fold symmetry that marks them as modes of the \LP\ system.  For each increase in $n$, modes are added in which the large amplitude oscillations are confined to triangle edges of level $n-2$.  Figure~\ref{fig:panel_of_modes} shows examples of such modes, along with additional modes in which two levels are excited.  For each mode in which the excitations are confined to level $n$ (or a set of levels up to $n$), there are corresponding modes confined to level $n+1$ (or a set up to $n+1$).  All modes within these hierarchies are high frequency modes in which neighboring masses on every level-$n$ edge oscillate out of phase with each other, as indicated by the black arrows in the first column of Fig.~\ref{fig:panel_of_modes}.  The modes in the first two rows of Fig.~\ref{fig:panel_of_modes} are 3-fold symmetric, while those in the bottom row are 6-fold symmetric. In the 6-fold symmetric modes, the instantaneous pattern around each triangle is chiral and every triangle has the same chirality. Note that the participation ratios in a given hierarchy decrease dramatically (roughly, by a factor of 2) with increasing $n$.

We now focus in more detail on the set of modes with the simplest geometry, those with the form of Fig.~\ref{fig:panel_of_modes}(c1).
(The modes in Fig.~\ref{fig:panel_of_modes}(b) actually have lower participation ratios, but those modes are degenerate and therefore less straightforward to analyze.)
We refer to a mode within this hierarchy as a level-$n$ edge mode and denote its frequency by $\omega_n$. To extract modes of this type from the computed spectrum, we construct a template that captures the essential structure of the mode and search for modes that have a high overlap with the template. 

\begin{figure}
  \centering
  \includegraphics[width=0.8\columnwidth]{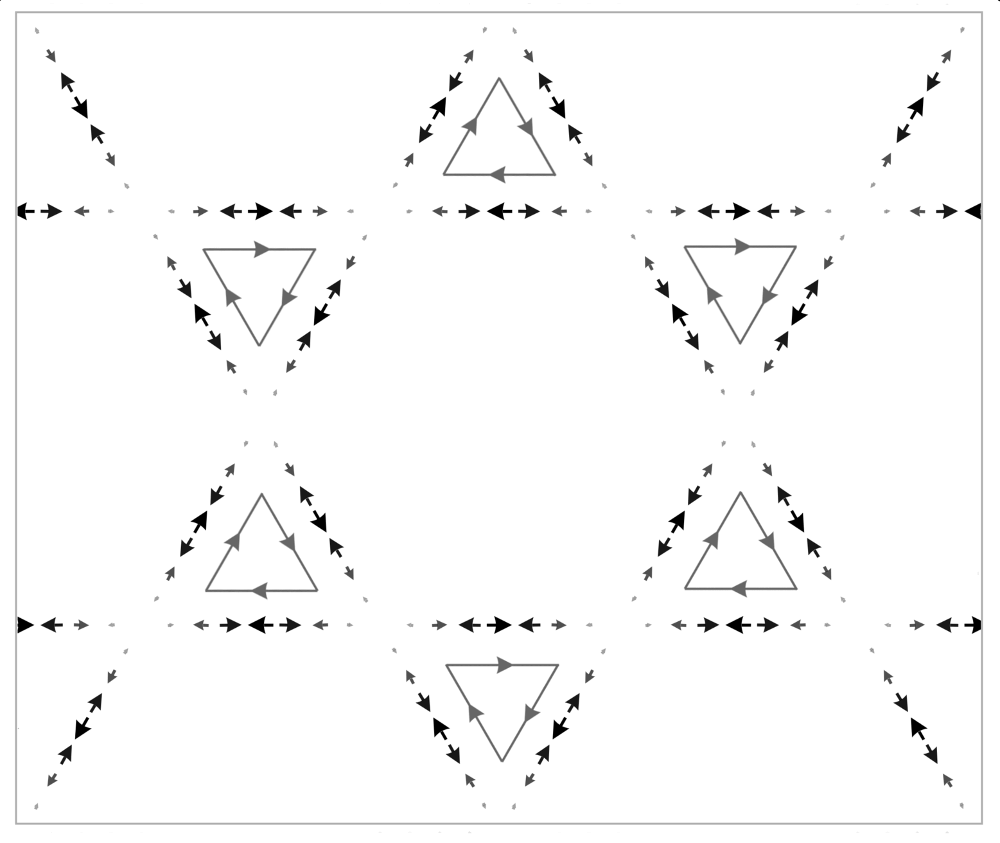}
  \caption{ A section of the $n=4$ template used to identify the edge mode of Fig.~\ref{fig:panel_of_modes}(c1). Black arrows represent the displacement vectors at a given time. Gray arrows illustrate that in every triangle, the instantaneous chiralities associated with the motion of the particles at edge centers are the same.}
  \label{fig:masks}
\end{figure}

For the level-$n$ template embedded in a $m$-periodic approximant with $m\ge n+2$, 
we assign a polarization vectors to each particle as follows.  Define lattice vectors 
\begin{equation}
\vec e_{\lambda} \equiv \l(\cos(2 \pi \lambda/3), \sin(2 \lambda\pi/3)\r)\,,
\end{equation} 
a normalization constant $c_{nm} \equiv 9\times 2^{2m-n-5}$, and the quantities $k_n \equiv 2^{n-1}$ and $z_n(i)\equiv i\!\! \mod\! 2^n$.
The polarization vector for the particle $i$ at position $\vec R = i_0 \vec e_0 + i_1 \vec e_1$ is given by
\begin{align}
\vec v_i^{(n)} = \bigg[ &(-1)^{i_0}\sin\l(\frac{i_0\pi}{k_n}\r)\delta(z_n(i_1),k_n)\,\vec e_0  \notag \\
- &(-1)^{i_1}\sin\l(\frac{i_1\pi}{k_n}\r)\delta(z_n(i_0),k_n)\,\vec e_1 \label{eq:template} \\
-&(-1)^{i_1}\sin\l(\frac{i_1\pi}{k_n}\r)\delta(z_n (i_0-i_1),k_n)\,\vec e_2 \bigg]\frac{1}{c_{nm}}\,, \notag
\end{align}
where $\delta(a,b)$ is the Kronecker delta. The normalization constant $c_{nm}$ is defined such that $\sum_{i} \vec v_i^{(n)} \cdot \vec v_i^{(n)} = 1$, where the sum runs over the sites in one unit cell of the $m$-periodic structure. Figure~\ref{fig:masks} shows a section of  the level-4 edge mode template.

To locate modes of interest, we scan through the high-frequency $\vec q = 0$ modes, calculating the overlap $I$ of the numerically calculated mode with $\vec v$:
\begin{equation}
I = \left|\sum_{i = 1}^N \vec v_i^{(n)}\cdot \boldsymbol \epsilon_i \right|\,.
\end{equation}
Modes with $I > 1/2$ are the relevant edge modes. Table~\ref{tab:mode_indices} gives the position of the mode in the list sorted from high to low frequency. All of the identified edge modes have frequencies that are within the highest 8\% of the relevant spectrum.
\begin{table}[hb]
	\centering
	\begin{ruledtabular}
	\begin{tabular}{ r  c  c  c  c}
				 & level-3 & level-4 & level-5 & level-6\\
	5-periodic & 29 & - &  - & - \\
	6-periodic & 119 & 27 &  - & - \\
	7-periodic & 482 & 108 &  22 & - \\
	8-periodic & 1916 & 432 &  82 & 22 \\
	\end{tabular}
	\end{ruledtabular}
	\caption{ The position in the spectrum of the edge modes of interest in four $n$-periodic structures.}
	\label{tab:mode_indices}
\end{table}


\section{Origin of the edge modes}
\label{sec:origin}

\begin{figure}
  \centering
  \includegraphics[width=0.45\textwidth]{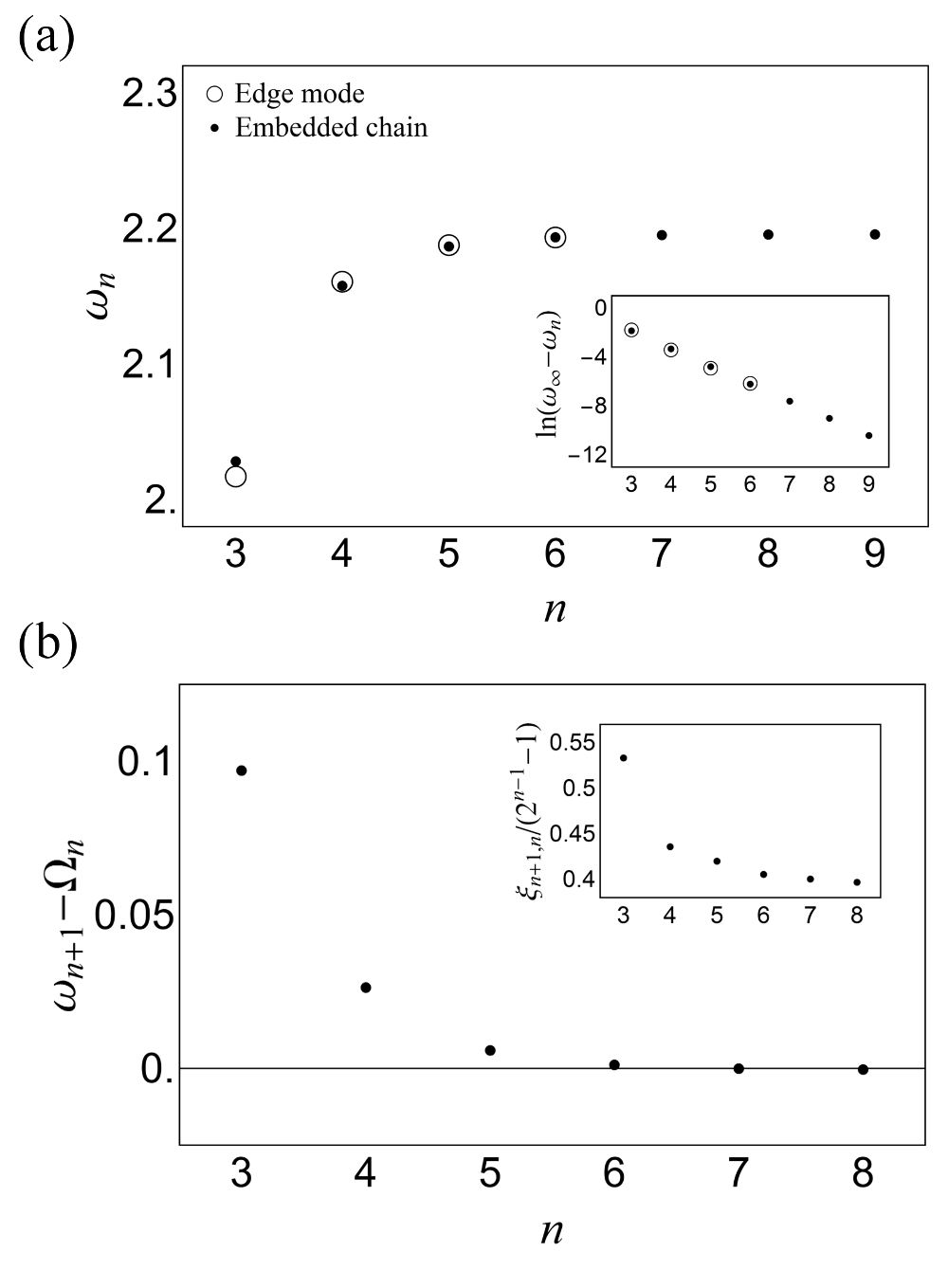}
  \caption{(a) Frequencies and estimated frequencies of the level-$n$ edge modes. Open circles are obtained using the methods described in Sec.~\ref{sec:computational_methods}. Closed circles are estimates based on a model of stiff chains embedded in a soft triangular lattice. The inset shows the same data on a log scale, where $\omega_{\infty}$ is the highest frequency that can propagate on the infinite embedded chain. (b) The difference between the estimated frequency of the level-$(n+1)$ edge mode and the highest frequency that can propagate on the level-$n$ chain. The inset shows the ratio of the decay length $\xi_{n+1,n}$ to the length of a level-$n$ edge $2^{n-1}-1$.  (See text.)
 }
  \label{fig:w_vs_n}
\end{figure}

The existence of a level-$n$ edge mode depends on the inability of lower level triangles to sustain oscillations at the necessary frequency.  That is, the frequency of the edge mode is higher than the highest frequency that can propagate through the lower level triangles that make up the bulk of the system.  Figure~\ref{fig:w_vs_n}(a) shows the numerically exact frequencies of the edge modes of different levels, along with calculated values based on the theory described below.

High amplitude oscillations do not occur on levels ($m<n$) in the level-$n$ edge modes because the lower levels cannot support propagation of a wave with frequency $\omega_n$.  The highest frequency mode that can propagate on the level-1 and level-2 structures is $\Omega_{1,2} = 1.90$, corresponding to the highest frequency mode of a kagome lattice with coupling strength $0.6k_{bb}$.  This implies that in the high-frequency modes of the \LP\ structure, the large amplitude oscillations will be confined to the stiff chains with lengths greater than one. One can see that oscillations at $\omega_n$ cannot propagate deeply into levels-$(3 \le m < n )$. When a chain is driven at one point at a frequency $\omega$ above the highest frequency in its spectrum, $\Omega$, the excitation will be localized with a decay length 
\begin{equation}
	\xi \approx \frac{1}{2\sqrt{\delta}}\,,
	\label{eq:decay}
\end{equation}
where $\delta = \omega - \Omega \ll 1$~\cite{adkins_JPhysCM2}. To estimate the decay length $\xi_{n,m}$ of the level-$n$ edge mode with frequency $\omega_n$ into level $m$ we take $\delta_{n,m} = \omega_n - \Omega_{m}$, where $\Omega_{m}$ is the highest frequency supported by the level-$m$ structure.

\begin{figure*}
  \centering
  \includegraphics[width=\textwidth]{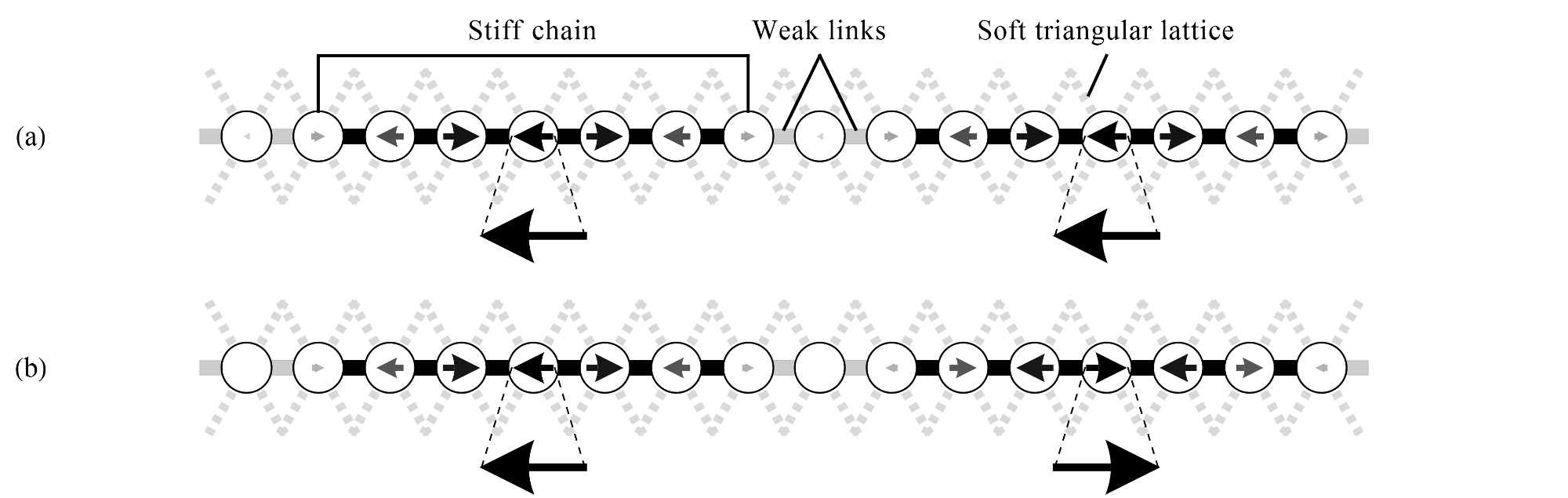}
  \caption{The level-4 embedded chain used to compute approximate frequencies.  (a) The mode corresponding to the highest frequency, $\Omega_4$.  (b) The mode corresponding to the level-4 edge mode with frequency $\omega_4$.  Similar chains are used to compute approximate values of $\omega_n$ and $\Omega_n$ for $3 \le n\le 9$.  The dashed bonds shown are a portion of a 2D triangular lattice.}
  \label{fig:approximating_chain}
\end{figure*}

\begin{figure*}
  \centering
   \includegraphics[width=\textwidth]{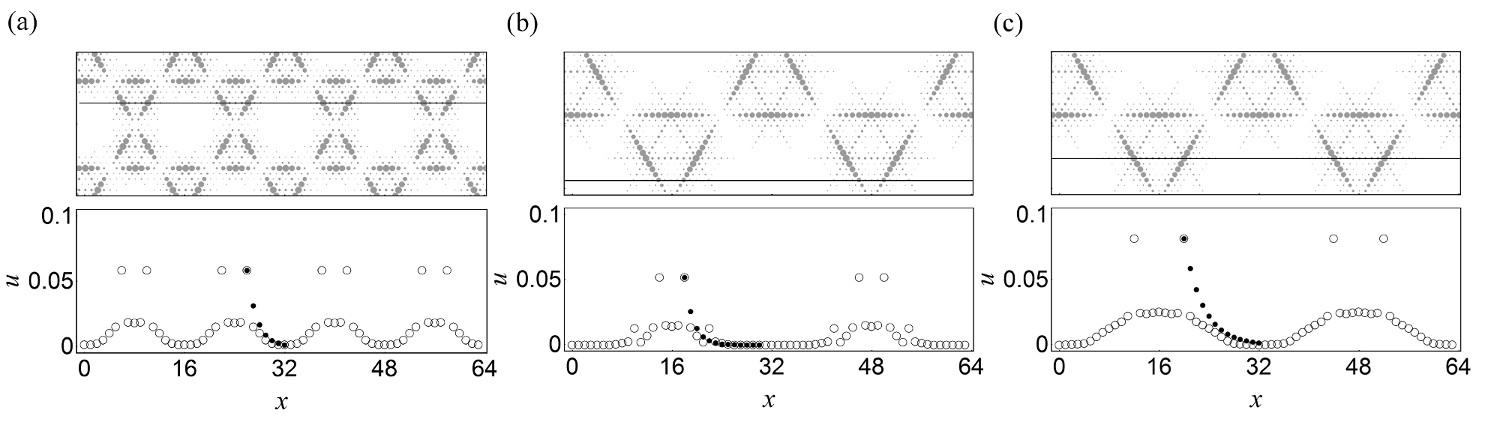}
  \caption{Decays of oscillations in edge modes.  (a) Level-3 edges in the level-4 edge mode. (b) Level-3 edges in the level-5 edge mode. (c) Level-4 edges in the level-5 edge mode. Open circles represent the magnitudes of the polarizations of the masses lying along the black line in the images above the plots. Filled circles are the amplitudes of an exponentially decaying function with decay length determined using Eq.~\eqref{eq:decay}.  }
  \label{fig:compare_exp}
\end{figure*}

We estimate $\Omega_{m}$ by calculating the phonon spectrum of an approximation to the level-$m$ structure: a line of stiff chain segments of length $2^{m-1}-1$ connected by pairs of weak links and embedded in a soft triangular lattice. (See Fig.~\ref{fig:approximating_chain}.) The stiff bonds were assigned a coupling strength $k_{bb}$ and the weak bonds were assigned a coupling strength $0.6 k_{bb}$.  The coupling strength for the bulk triangular lattice, $0.619 k_{bb}$, was chosen so that the frequency of the level-6 chain matches exact computation. Figure~\ref{fig:compare_exp} shows the predicted decay curves overlaid on the oscillation amplitudes of two modes obtained from the full phonon calculations for the 7-periodic structure. In all three plots, the predicted decay lengths account well for the numerically determined amplitudes.

One might worry that as $n$ increases, the decreasing frequency difference between adjacent levels may result in modes that have large amplitude oscillations on levels $(m<n)$. We estimate $\omega_n$ and $\Omega_{n}$ for levels larger than those we have numerically computed by again using the chain illustrated in Fig.~\ref{fig:approximating_chain}. The frequency $\omega_n$ is that of the mode with polarization vectors similar to the actual level-$n$ edge mode (Fig.~\ref{fig:approximating_chain}(b)). Although the difference between $\omega_{n+1}$ and $\Omega_n$ decreases quickly with $n$, as shown in Fig.~\ref{fig:w_vs_n}(b), the ratio of the decay length of the level-$n$ edge mode into the level-$(n-1)$ structure to the length of a level-$(n-1)$ triangle edge (Fig.~\ref{fig:w_vs_n}(b) inset) shows a trend towards lower values, a strong indication that the edge mode structure persists to arbitrarily large $n$. 

\begin{figure}
  \centering
   \includegraphics[width=0.8\columnwidth]{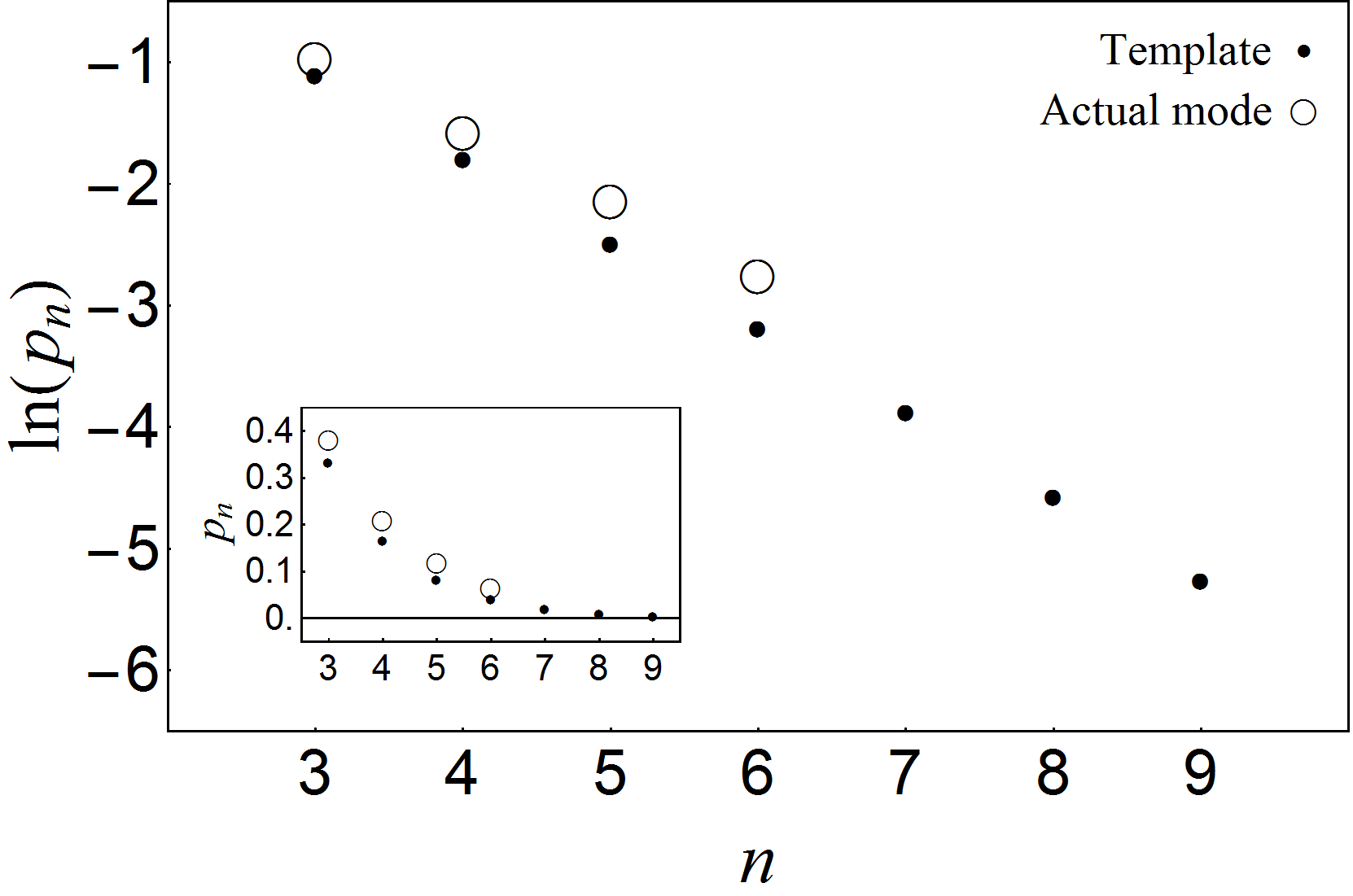}
  \caption{Participation ratios of level-$n$ edge modes (open circles) obtained as described in Sec.~\ref{sec:computational_methods} and the level-$n$ templates (filled circles) generated by Eq.~\eqref{eq:template}.}
  \label{fig:participation_ratios}
\end{figure}

We estimate the behavior of the participation ratio of the level-$n$ edge mode $p_n^{\rm{actual}}$ by determining the functional form of the participation ratio of the level-$n$ template $p_n^{\rm{temp}}$. From Eqs.~\eqref{eq:participation_ratio}~and~\eqref{eq:template} a straightforward derivation gives
\begin{equation}
p_n^{\rm{temp}} = \frac{1}{3\times2^{n-3}}\,,
\end{equation}
which decreases by a factor of two with each level. The actual level-$n$ edge modes deviate from the template due to the exponential decay into the bulk. The participation ratios of both the templates and the actual modes are presented in Fig.~\ref{fig:participation_ratios} for multiple $n$. Although the slope of $p_n^{\rm{actual}}(n)$ up to $n=6$ appears smaller than the expected scaling, we conjecture that the scaling will recover at larger $n$. Numerical confirmation of the $2^{-n}$ scaling would require greater computational capacity. 

To support the claim that the level-$n$ edge modes of the 7-periodic structure exist within the spectrum of the \LP\ structure, we verify the persistence of the modes as the size of the unit cell is changed.  We find that for the 5-, 6-, 7-, and 8-periodic approximants, the frequencies of a level-$n$ edge mode that exists in more than one approximant are the same up to five significant digits.

\begin{figure}
  \centering
  \includegraphics[width=0.4\textwidth]{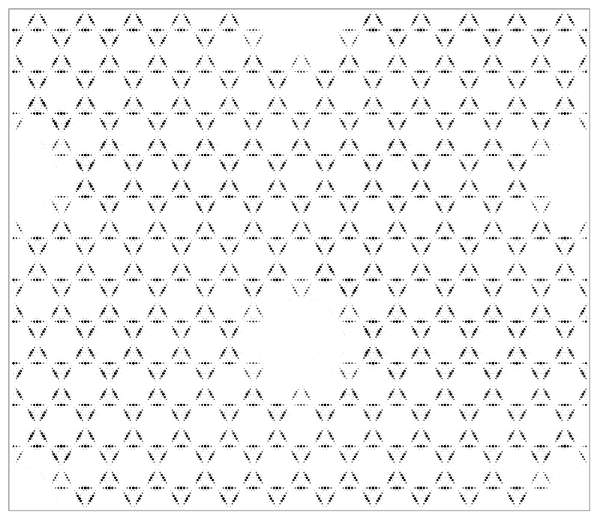}
  \caption{The level-4 edge mode of the 8-periodic structure when a single mass along and near the center of a level-4 edge is removed. Multiple unit cells are shown; the removal of one mass along a level-4 edge in an infinite version of the structure would create a single hole. }
  \label{fig:level-8_vacancy}
\end{figure}

Because the frequencies of edge modes lie outside the spectrum of the bulk comprised of lower level triangles, we expect the modes to be robust to some degree of disorder.  Figure~\ref{fig:level-8_vacancy} shows a level-$n$ mode in a system where a mass that would oscillate at high amplitude is removed. The vacancy gives rise to a hole in the pattern, but the long-range order of the mode persists.

Introducing disorder into the coupling strengths destroys the long-range order of the mode but does not increase the participation ratio. We calculated the phonon modes of the 7-periodic structure with each coupling strength multiplied by a random number between $1-\alpha$ and $1+\alpha$. We find that the high-frequency modes remain localized along the stiff chains for all values of $\alpha$ used, but even for $\alpha$ as small as 0.01, the mode can no longer be identified using the template of Eq.~\eqref{eq:template}.  As expected, disorder results in localization along the 1D chains, resulting in even lower participation ratios~\cite{dean_PPS84}.

\section{Conclusion}

We have shown that a \LP\ ball and spring model supports modes with arbitrarily small participation ratios.  These are not exponentially localized modes, but instead are extended modes in which the large amplitude oscillations are confined to sparse periodic nets. Two properties of the \LP\ structure enable it to support modes with arbitrarily low participation ratios. First, the presence of stiff chains  embedded in a softer bulk allows for high frequency modes confined to those chains.  The exclusion of the oscillations from the bulk also results in the confinement of the modes being robust to vacancies and some degree of disorder in the spring constants.  Second, the \LP\ system is comprised of a hierarchy of increasingly stiff and sparse networks of chains.  At each level, the chains are stiff enough to support modes with sufficiently short decay lengths in the bulk that the confinement is sharply defined.  The result is that for an arbitrarily small choice of $p$, there exists a level in the hierarchy that supports modes with participation ratios less than $p$.

Questions remain about the nature of the \LP\ spectrum. We have  studied in depth only a subset of the low participation ratio modes, in particular, the level-$n$ edge modes.  Though we have not observed any obvious structural features of the spectrum, a closer look is likely to reveal nontrivial scaling laws.  We have also not studied our model in the regime where $\alpha > 1$, in which case the longer triangle edges cannot support the highest frequency modes.  Most importantly, a more realistic model of a colloidal phase formed from structured particles will have to include the degrees of freedom associated with rotations of the tiles.  We conjecture that modes confined to level-$n$ sublattices will still be generic in parameter regimes where bonds associated with triangle corners are weaker than bonds associated with triangle edges.

\begin{acknowledgments}
Support for this research was provided by the NSF's Research Triangle MRSEC (DMR-1121107).  
\end{acknowledgments}

\appendix*
\section{Pattern of spring stiffnesses for the \LP\ ball and spring model}

Here we give a precise description of the pattern of coupling strengths in a \LP\ ball-and-spring model, corresponding to the structure in Fig.~\ref{fig:5-periodic_removed}(c). 

Define the unit vectors 
\begin{equation}
\vec e_{\lambda} = \l(\cos(2 \pi \lambda/3), \sin(2 \pi \lambda/3)\r)\,, 
\end{equation}
with $\lambda \in \{0,1,2\}$.
The point masses lie on lattice sites $i_0 \vec e_0 + i_1 \vec e_1$, with $i_0,i_1 \in \mathbb{Z}$. 
Consider the bond joining sites $\vec r$ and $\vec s =\vec r + \vec e_{\lambda}$. Assign to this bond a pair of integers
\begin{equation}
	(i,j) = \frac{2}{\sqrt{3}}\l( \vec r \cdot \vec e'_{\lambda+1}\,, -\vec r \cdot \vec e'_{\lambda+2}\r)\,,
\end{equation}
where subscripts are taken modulo 3 and 
$\vec e'_{\lambda}$ is a rotation of $\vec e_{\lambda}$ by $\pi/2$.

Let $\mathrm{GCD}(a,b)$ be the greatest common divisor of $a$ and $b$, and define $Q(n) \equiv \mathrm{GCD}(2^n, n)$, taking $Q(0) \equiv \infty$.  For $\vec r, \vec s \ne (0,0)$, the stiffness $k$ is given by 
\begin{equation}
k = \left\{ \begin{array}{ll}
 k_{bb} & \rm{if\ } Q(|i|) = Q(|j|) \wedge Q(|i - 1|) = Q(|j - 1|) \\
 \alpha k_{bb} & \rm{otherwise}\,.
\end{array} \right.
\end{equation}

For bonds connecting to (0,0), if the bond is in the $\pm \vec{e}_0$ direction, the stiffness is $k_{bb}$, otherwise it is $\alpha k_{bb}$.


\begin{thebibliography}{26}%
\makeatletter
\providecommand \@ifxundefined [1]{%
 \@ifx{#1\undefined}
}%
\providecommand \@ifnum [1]{%
 \ifnum #1\expandafter \@firstoftwo
 \else \expandafter \@secondoftwo
 \fi
}%
\providecommand \@ifx [1]{%
 \ifx #1\expandafter \@firstoftwo
 \else \expandafter \@secondoftwo
 \fi
}%
\providecommand \natexlab [1]{#1}%
\providecommand \enquote  [1]{``#1''}%
\providecommand \bibnamefont  [1]{#1}%
\providecommand \bibfnamefont [1]{#1}%
\providecommand \citenamefont [1]{#1}%
\providecommand \href@noop [0]{\@secondoftwo}%
\providecommand \href [0]{\begingroup \@sanitize@url \@href}%
\providecommand \@href[1]{\@@startlink{#1}\@@href}%
\providecommand \@@href[1]{\endgroup#1\@@endlink}%
\providecommand \@sanitize@url [0]{\catcode `\\12\catcode `\$12\catcode
  `\&12\catcode `\#12\catcode `\^12\catcode `\_12\catcode `\%12\relax}%
\providecommand \@@startlink[1]{}%
\providecommand \@@endlink[0]{}%
\providecommand \url  [0]{\begingroup\@sanitize@url \@url }%
\providecommand \@url [1]{\endgroup\@href {#1}{\urlprefix }}%
\providecommand \urlprefix  [0]{URL }%
\providecommand \Eprint [0]{\href }%
\providecommand \doibase [0]{http://dx.doi.org/}%
\providecommand \selectlanguage [0]{\@gobble}%
\providecommand \bibinfo  [0]{\@secondoftwo}%
\providecommand \bibfield  [0]{\@secondoftwo}%
\providecommand \translation [1]{[#1]}%
\providecommand \BibitemOpen [0]{}%
\providecommand \bibitemStop [0]{}%
\providecommand \bibitemNoStop [0]{.\EOS\space}%
\providecommand \EOS [0]{\spacefactor3000\relax}%
\providecommand \BibitemShut  [1]{\csname bibitem#1\endcsname}%
\let\auto@bib@innerbib\@empty
\bibitem [{\citenamefont {Barker~Jr.}\ and\ \citenamefont
  {Sievers}(1975)}]{sievers_RMP47}%
  \BibitemOpen
  \bibfield  {author} {\bibinfo {author} {\bibfnamefont {A.~S.}\ \bibnamefont
  {Barker~Jr.}}\ and\ \bibinfo {author} {\bibfnamefont {A.~J.}\ \bibnamefont
  {Sievers}},\ }\href@noop {} {\bibfield  {journal} {\bibinfo  {journal}
  {Reviews of Modern Physics}\ }\textbf {\bibinfo {volume} {47}},\ \bibinfo
  {pages} {2} (\bibinfo {year} {1975})}\BibitemShut {NoStop}%
\bibitem [{\citenamefont {D{\"u}ring}\ \emph {et~al.}(2013)\citenamefont
  {D{\"u}ring}, \citenamefont {Lerner},\ and\ \citenamefont
  {Wyart}}]{during_SoftMatt9}%
  \BibitemOpen
  \bibfield  {author} {\bibinfo {author} {\bibfnamefont {G.}~\bibnamefont
  {D{\"u}ring}}, \bibinfo {author} {\bibfnamefont {E.}~\bibnamefont {Lerner}},
  \ and\ \bibinfo {author} {\bibfnamefont {M.}~\bibnamefont {Wyart}},\
  }\href@noop {} {\bibfield  {journal} {\bibinfo  {journal} {Soft Matter}\
  }\textbf {\bibinfo {volume} {9}},\ \bibinfo {pages} {146} (\bibinfo {year}
  {2013})}\BibitemShut {NoStop}%
\bibitem [{\citenamefont {Quilichini}\ and\ \citenamefont
  {Janssen}(1997)}]{janssen_RMP69}%
  \BibitemOpen
  \bibfield  {author} {\bibinfo {author} {\bibfnamefont {M.}~\bibnamefont
  {Quilichini}}\ and\ \bibinfo {author} {\bibfnamefont {T.}~\bibnamefont
  {Janssen}},\ }\href@noop {} {\bibfield  {journal} {\bibinfo  {journal}
  {Reviews of Modern Physics}\ }\textbf {\bibinfo {volume} {69}},\ \bibinfo
  {pages} {1} (\bibinfo {year} {1997})}\BibitemShut {NoStop}%
\bibitem [{\citenamefont {Sun}\ \emph {et~al.}(2012)\citenamefont {Sun},
  \citenamefont {Souslov}, \citenamefont {Mao},\ and\ \citenamefont
  {Lubensky}}]{lubensky_PNAS109}%
  \BibitemOpen
  \bibfield  {author} {\bibinfo {author} {\bibfnamefont {K.}~\bibnamefont
  {Sun}}, \bibinfo {author} {\bibfnamefont {A.}~\bibnamefont {Souslov}},
  \bibinfo {author} {\bibfnamefont {X.}~\bibnamefont {Mao}}, \ and\ \bibinfo
  {author} {\bibfnamefont {T.~C.}\ \bibnamefont {Lubensky}},\ }\href@noop {}
  {\bibfield  {journal} {\bibinfo  {journal} {Proceedings of the National
  Academy of Sciences}\ }\textbf {\bibinfo {volume} {109}},\ \bibinfo {pages}
  {31} (\bibinfo {year} {2012})}\BibitemShut {NoStop}%
\bibitem [{\citenamefont {Paulose}\ \emph {et~al.}(2015)\citenamefont
  {Paulose}, \citenamefont {Chen},\ and\ \citenamefont
  {Vitelli}}]{vitelli_NatPhys11}%
  \BibitemOpen
  \bibfield  {author} {\bibinfo {author} {\bibfnamefont {J.}~\bibnamefont
  {Paulose}}, \bibinfo {author} {\bibfnamefont {B.~G.}\ \bibnamefont {Chen}}, \
  and\ \bibinfo {author} {\bibfnamefont {V.}~\bibnamefont {Vitelli}},\
  }\href@noop {} {\bibfield  {journal} {\bibinfo  {journal} {Nature Physics}\
  }\textbf {\bibinfo {volume} {11}},\ \bibinfo {pages} {153} (\bibinfo {year}
  {2015})}\BibitemShut {NoStop}%
\bibitem [{\citenamefont {Prodan}\ and\ \citenamefont
  {Prodan}(2009)}]{prodan_PRL103}%
  \BibitemOpen
  \bibfield  {author} {\bibinfo {author} {\bibfnamefont {E.}~\bibnamefont
  {Prodan}}\ and\ \bibinfo {author} {\bibfnamefont {C.}~\bibnamefont
  {Prodan}},\ }\href@noop {} {\bibfield  {journal} {\bibinfo  {journal}
  {Physical Review Letters}\ }\textbf {\bibinfo {volume} {103}},\ \bibinfo
  {pages} {248101} (\bibinfo {year} {2009})}\BibitemShut {NoStop}%
\bibitem [{\citenamefont {Pal}\ \emph {et~al.}(2015)\citenamefont {Pal},
  \citenamefont {Shaeffer},\ and\ \citenamefont
  {Ruzzene}}]{pal_arXiv:1511.07507}%
  \BibitemOpen
  \bibfield  {author} {\bibinfo {author} {\bibfnamefont {R.~K.}\ \bibnamefont
  {Pal}}, \bibinfo {author} {\bibfnamefont {M.}~\bibnamefont {Shaeffer}}, \
  and\ \bibinfo {author} {\bibfnamefont {M.}~\bibnamefont {Ruzzene}},\
  }\href@noop {} {\bibfield  {journal} {\bibinfo  {journal} {arXiv}\ }\textbf
  {\bibinfo {volume} {1511}},\ \bibinfo {pages} {07507} (\bibinfo {year}
  {2015})}\BibitemShut {NoStop}%
\bibitem [{\citenamefont {Ambati}\ \emph {et~al.}(2007)\citenamefont {Ambati},
  \citenamefont {Fang}, \citenamefont {Sun},\ and\ \citenamefont
  {Zhang}}]{zhang_PRB75}%
  \BibitemOpen
  \bibfield  {author} {\bibinfo {author} {\bibfnamefont {M.}~\bibnamefont
  {Ambati}}, \bibinfo {author} {\bibfnamefont {N.}~\bibnamefont {Fang}},
  \bibinfo {author} {\bibfnamefont {C.}~\bibnamefont {Sun}}, \ and\ \bibinfo
  {author} {\bibfnamefont {X.}~\bibnamefont {Zhang}},\ }\href@noop {}
  {\bibfield  {journal} {\bibinfo  {journal} {Physical Review B}\ }\textbf
  {\bibinfo {volume} {75}},\ \bibinfo {pages} {195447} (\bibinfo {year}
  {2007})}\BibitemShut {NoStop}%
\bibitem [{\citenamefont {Wu}\ \emph {et~al.}(2009)\citenamefont {Wu},
  \citenamefont {Chen},\ and\ \citenamefont {Liu}}]{liu_APL95}%
  \BibitemOpen
  \bibfield  {author} {\bibinfo {author} {\bibfnamefont {L.-Y.}\ \bibnamefont
  {Wu}}, \bibinfo {author} {\bibfnamefont {L.-W.}\ \bibnamefont {Chen}}, \ and\
  \bibinfo {author} {\bibfnamefont {C.-M.}\ \bibnamefont {Liu}},\ }\href@noop
  {} {\bibfield  {journal} {\bibinfo  {journal} {Applied Physics Letters}\
  }\textbf {\bibinfo {volume} {95}},\ \bibinfo {pages} {013506} (\bibinfo
  {year} {2009})}\BibitemShut {NoStop}%
\bibitem [{\citenamefont {Khelif}\ \emph {et~al.}(2004)\citenamefont {Khelif},
  \citenamefont {Choujaa}, \citenamefont {Benchabane}, \citenamefont
  {Djafari-Rouhani},\ and\ \citenamefont {Laude}}]{laude_APL84}%
  \BibitemOpen
  \bibfield  {author} {\bibinfo {author} {\bibfnamefont {A.}~\bibnamefont
  {Khelif}}, \bibinfo {author} {\bibfnamefont {A.}~\bibnamefont {Choujaa}},
  \bibinfo {author} {\bibfnamefont {S.}~\bibnamefont {Benchabane}}, \bibinfo
  {author} {\bibfnamefont {B.}~\bibnamefont {Djafari-Rouhani}}, \ and\ \bibinfo
  {author} {\bibfnamefont {V.}~\bibnamefont {Laude}},\ }\href@noop {}
  {\bibfield  {journal} {\bibinfo  {journal} {Applied Physics Letters}\
  }\textbf {\bibinfo {volume} {84}},\ \bibinfo {pages} {4400} (\bibinfo {year}
  {2004})}\BibitemShut {NoStop}%
\bibitem [{\citenamefont {Witten}(2007)}]{witten_RMP79}%
  \BibitemOpen
  \bibfield  {author} {\bibinfo {author} {\bibfnamefont {T.}~\bibnamefont
  {Witten}},\ }\href@noop {} {\bibfield  {journal} {\bibinfo  {journal}
  {Reviews of Modern Physics}\ }\textbf {\bibinfo {volume} {79}},\ \bibinfo
  {pages} {2} (\bibinfo {year} {2007})}\BibitemShut {NoStop}%
\bibitem [{\citenamefont {Chen}\ \emph {et~al.}(2013)\citenamefont {Chen},
  \citenamefont {Still}, \citenamefont {Shoenholz}, \citenamefont {Aptowicz},
  \citenamefont {Schindler}, \citenamefont {Maggs}, \citenamefont {Liu},\ and\
  \citenamefont {Yodh}}]{yodh_PRE88}%
  \BibitemOpen
  \bibfield  {author} {\bibinfo {author} {\bibfnamefont {K.}~\bibnamefont
  {Chen}}, \bibinfo {author} {\bibfnamefont {T.}~\bibnamefont {Still}},
  \bibinfo {author} {\bibfnamefont {S.}~\bibnamefont {Shoenholz}}, \bibinfo
  {author} {\bibfnamefont {K.~B.}\ \bibnamefont {Aptowicz}}, \bibinfo {author}
  {\bibfnamefont {M.}~\bibnamefont {Schindler}}, \bibinfo {author}
  {\bibfnamefont {A.~C.}\ \bibnamefont {Maggs}}, \bibinfo {author}
  {\bibfnamefont {A.~J.}\ \bibnamefont {Liu}}, \ and\ \bibinfo {author}
  {\bibfnamefont {A.~G.}\ \bibnamefont {Yodh}},\ }\href@noop {} {\bibfield
  {journal} {\bibinfo  {journal} {Physical Review E}\ }\textbf {\bibinfo
  {volume} {88}},\ \bibinfo {pages} {022315} (\bibinfo {year}
  {2013})}\BibitemShut {NoStop}%
\bibitem [{\citenamefont {Sigalas}(1998)}]{sigalas_JAP84}%
  \BibitemOpen
  \bibfield  {author} {\bibinfo {author} {\bibfnamefont {M.~M.}\ \bibnamefont
  {Sigalas}},\ }\href@noop {} {\bibfield  {journal} {\bibinfo  {journal}
  {Journal of Applied Physics}\ }\textbf {\bibinfo {volume} {84}},\ \bibinfo
  {pages} {3026} (\bibinfo {year} {1998})}\BibitemShut {NoStop}%
\bibitem [{\citenamefont {Socolar}\ and\ \citenamefont
  {Taylor}(2011)}]{socolar-taylor_JCombTh11}%
  \BibitemOpen
  \bibfield  {author} {\bibinfo {author} {\bibfnamefont {J.~E.~S.}\
  \bibnamefont {Socolar}}\ and\ \bibinfo {author} {\bibfnamefont {J.~M.}\
  \bibnamefont {Taylor}},\ }\href@noop {} {\bibfield  {journal} {\bibinfo
  {journal} {Journal of Combinatorial Theory, Series A}\ }\textbf {\bibinfo
  {volume} {118}},\ \bibinfo {pages} {2207} (\bibinfo {year}
  {2011})}\BibitemShut {NoStop}%
\bibitem [{\citenamefont {Marcoux}\ \emph {et~al.}(2014)\citenamefont
  {Marcoux}, \citenamefont {Byington}, \citenamefont {Qian}, \citenamefont
  {Charbonneau},\ and\ \citenamefont {Socolar}}]{marcoux-socolar_PRE90}%
  \BibitemOpen
  \bibfield  {author} {\bibinfo {author} {\bibfnamefont {C.}~\bibnamefont
  {Marcoux}}, \bibinfo {author} {\bibfnamefont {T.~W.}\ \bibnamefont
  {Byington}}, \bibinfo {author} {\bibfnamefont {Z.}~\bibnamefont {Qian}},
  \bibinfo {author} {\bibfnamefont {P.}~\bibnamefont {Charbonneau}}, \ and\
  \bibinfo {author} {\bibfnamefont {J.~E.~S.}\ \bibnamefont {Socolar}},\
  }\href@noop {} {\bibfield  {journal} {\bibinfo  {journal} {Physical Review
  E}\ }\textbf {\bibinfo {volume} {90}},\ \bibinfo {pages} {012136} (\bibinfo
  {year} {2014})}\BibitemShut {NoStop}%
\bibitem [{\citenamefont {Byington}\ and\ \citenamefont
  {Socolar}(2012)}]{byington-socolar_PRL108}%
  \BibitemOpen
  \bibfield  {author} {\bibinfo {author} {\bibfnamefont {T.~W.}\ \bibnamefont
  {Byington}}\ and\ \bibinfo {author} {\bibfnamefont {J.~E.~S.}\ \bibnamefont
  {Socolar}},\ }\href@noop {} {\bibfield  {journal} {\bibinfo  {journal}
  {Physical Review Letters}\ }\textbf {\bibinfo {volume} {108}},\ \bibinfo
  {pages} {045701} (\bibinfo {year} {2012})}\BibitemShut {NoStop}%
\bibitem [{\citenamefont {Wang}\ \emph {et~al.}(2012)\citenamefont {Wang},
  \citenamefont {Wang}, \citenamefont {Breed}, \citenamefont {Manoharan},
  \citenamefont {Feng}, \citenamefont {Hollingsworth}, \citenamefont {Weck},\
  and\ \citenamefont {Pine}}]{wang-pine_Nature}%
  \BibitemOpen
  \bibfield  {author} {\bibinfo {author} {\bibfnamefont {Y.}~\bibnamefont
  {Wang}}, \bibinfo {author} {\bibfnamefont {Y.}~\bibnamefont {Wang}}, \bibinfo
  {author} {\bibfnamefont {D.~R.}\ \bibnamefont {Breed}}, \bibinfo {author}
  {\bibfnamefont {W.~N.}\ \bibnamefont {Manoharan}}, \bibinfo {author}
  {\bibfnamefont {L.}~\bibnamefont {Feng}}, \bibinfo {author} {\bibfnamefont
  {A.~D.}\ \bibnamefont {Hollingsworth}}, \bibinfo {author} {\bibfnamefont
  {M.}~\bibnamefont {Weck}}, \ and\ \bibinfo {author} {\bibfnamefont {D.~J.}\
  \bibnamefont {Pine}},\ }\href@noop {} {\bibfield  {journal} {\bibinfo
  {journal} {Nature}\ }\textbf {\bibinfo {volume} {491}},\ \bibinfo {pages}
  {51} (\bibinfo {year} {2012})}\BibitemShut {NoStop}%
\bibitem [{\citenamefont {Yi}\ \emph {et~al.}(2013)\citenamefont {Yi},
  \citenamefont {Pine},\ and\ \citenamefont {Sacanna}}]{yi-sacanna_JPhysCM}%
  \BibitemOpen
  \bibfield  {author} {\bibinfo {author} {\bibfnamefont {G.-R.}\ \bibnamefont
  {Yi}}, \bibinfo {author} {\bibfnamefont {D.~J.}\ \bibnamefont {Pine}}, \ and\
  \bibinfo {author} {\bibfnamefont {S.}~\bibnamefont {Sacanna}},\ }\href@noop
  {} {\bibfield  {journal} {\bibinfo  {journal} {Journal of Physics: Condensed
  Matter}\ }\textbf {\bibinfo {volume} {25}},\ \bibinfo {pages} {193101}
  (\bibinfo {year} {2013})}\BibitemShut {NoStop}%
\bibitem [{\citenamefont {Feng}\ \emph {et~al.}(2013)\citenamefont {Feng},
  \citenamefont {Dreyfus}, \citenamefont {Sha}, \citenamefont {Seeman},\ and\
  \citenamefont {Chaikin}}]{feng-chaikin_AdvMat}%
  \BibitemOpen
  \bibfield  {author} {\bibinfo {author} {\bibfnamefont {L.}~\bibnamefont
  {Feng}}, \bibinfo {author} {\bibfnamefont {R.}~\bibnamefont {Dreyfus}},
  \bibinfo {author} {\bibfnamefont {R.}~\bibnamefont {Sha}}, \bibinfo {author}
  {\bibfnamefont {N.~C.}\ \bibnamefont {Seeman}}, \ and\ \bibinfo {author}
  {\bibfnamefont {P.~M.}\ \bibnamefont {Chaikin}},\ }\href@noop {} {\bibfield
  {journal} {\bibinfo  {journal} {Advanced Materials}\ }\textbf {\bibinfo
  {volume} {25}},\ \bibinfo {pages} {2779} (\bibinfo {year}
  {2013})}\BibitemShut {NoStop}%
\bibitem [{\citenamefont {Fleharty}\ \emph {et~al.}(2014)\citenamefont
  {Fleharty}, \citenamefont {van Swol},\ and\ \citenamefont
  {Petsev}}]{petsev_PRL113}%
  \BibitemOpen
  \bibfield  {author} {\bibinfo {author} {\bibfnamefont {M.~E.}\ \bibnamefont
  {Fleharty}}, \bibinfo {author} {\bibfnamefont {F.}~\bibnamefont {van Swol}},
  \ and\ \bibinfo {author} {\bibfnamefont {D.~N.}\ \bibnamefont {Petsev}},\
  }\href@noop {} {\bibfield  {journal} {\bibinfo  {journal} {Physical Review
  Letters}\ }\textbf {\bibinfo {volume} {113}},\ \bibinfo {pages} {158302}
  (\bibinfo {year} {2014})}\BibitemShut {NoStop}%
\bibitem [{\citenamefont {Socolar}\ and\ \citenamefont
  {Taylor}(2012)}]{socolar-tayor_MathInt12}%
  \BibitemOpen
  \bibfield  {author} {\bibinfo {author} {\bibfnamefont {J.~E.~S.}\
  \bibnamefont {Socolar}}\ and\ \bibinfo {author} {\bibfnamefont {J.~M.}\
  \bibnamefont {Taylor}},\ }\href@noop {} {\bibfield  {journal} {\bibinfo
  {journal} {The Mathematical Intelligencer}\ }\textbf {\bibinfo {volume}
  {34}},\ \bibinfo {pages} {1} (\bibinfo {year} {2012})}\BibitemShut {NoStop}%
\bibitem [{\citenamefont {Ashcroft}\ and\ \citenamefont
  {Mermin}(1976)}]{ashcroft_mermin_solidstate76}%
  \BibitemOpen
  \bibfield  {author} {\bibinfo {author} {\bibfnamefont {N.~W.}\ \bibnamefont
  {Ashcroft}}\ and\ \bibinfo {author} {\bibfnamefont {N.~D.}\ \bibnamefont
  {Mermin}},\ }\href@noop {} {\emph {\bibinfo {title} {Solid State Physics}}}\
  (\bibinfo  {publisher} {Harcourt College Publishers},\ \bibinfo {address}
  {New York},\ \bibinfo {year} {1976})\BibitemShut {NoStop}%
\bibitem [{\citenamefont {Bell}\ \emph {et~al.}(1970)\citenamefont {Bell},
  \citenamefont {Dean},\ and\ \citenamefont
  {Hibbins-Butler}}]{hibbins_butler_JPhysC1970}%
  \BibitemOpen
  \bibfield  {author} {\bibinfo {author} {\bibfnamefont {R.~J.}\ \bibnamefont
  {Bell}}, \bibinfo {author} {\bibfnamefont {P.}~\bibnamefont {Dean}}, \ and\
  \bibinfo {author} {\bibfnamefont {D.~C.}\ \bibnamefont {Hibbins-Butler}},\
  }\href@noop {} {\bibfield  {journal} {\bibinfo  {journal} {Journal of Physics
  C: Solid State Physics}\ }\textbf {\bibinfo {volume} {3}},\ \bibinfo {pages}
  {2111} (\bibinfo {year} {1970})}\BibitemShut {NoStop}%
\bibitem [{\citenamefont {Laird}\ and\ \citenamefont
  {Schober}(1991)}]{schober_PRL66}%
  \BibitemOpen
  \bibfield  {author} {\bibinfo {author} {\bibfnamefont {B.~B.}\ \bibnamefont
  {Laird}}\ and\ \bibinfo {author} {\bibfnamefont {H.~R.}\ \bibnamefont
  {Schober}},\ }\href@noop {} {\bibfield  {journal} {\bibinfo  {journal}
  {Physical Review Letters}\ }\textbf {\bibinfo {volume} {66}},\ \bibinfo
  {pages} {635} (\bibinfo {year} {1991})}\BibitemShut {NoStop}%
\bibitem [{\citenamefont {Adkins}(1990)}]{adkins_JPhysCM2}%
  \BibitemOpen
  \bibfield  {author} {\bibinfo {author} {\bibfnamefont {C.~J.}\ \bibnamefont
  {Adkins}},\ }\href@noop {} {\bibfield  {journal} {\bibinfo  {journal}
  {Journal of Physics: Condensed Matter}\ }\textbf {\bibinfo {volume} {2}},\
  \bibinfo {pages} {1445} (\bibinfo {year} {1990})}\BibitemShut {NoStop}%
\bibitem [{\citenamefont {Dean}(1964)}]{dean_PPS84}%
  \BibitemOpen
  \bibfield  {author} {\bibinfo {author} {\bibfnamefont {P.}~\bibnamefont
  {Dean}},\ }\href@noop {} {\bibfield  {journal} {\bibinfo  {journal}
  {Proceedings of the Physical Society}\ }\textbf {\bibinfo {volume} {84}},\
  \bibinfo {pages} {727} (\bibinfo {year} {1964})}\BibitemShut {NoStop}%
\end{thebibliography}
\end{document}